\documentclass[a4paper,fleqn,usenatbib]{mnras}
\usepackage{latexsym}
\usepackage{dcolumn}% Align table columns on decimal point
\usepackage{bm}% bold math
\usepackage{amsmath}
\usepackage{natbib}
\input{epsf}
\newcommand{\be}{\begin{equation}}
\newcommand{\ee}{\end{equation}}
\newcommand{\ba}{\begin{eqnarray}}
\newcommand{\ea}{\end{eqnarray}}
\newcommand{\bi}{\begin{itemize}}
\newcommand{\ei}{\end{itemize}}
\newcommand{\bfi}{\begin{figure}
\epsfxsize=9cm
\epsffile}
\newcommand{\bfig}{\begin{figure*}
\epsfxsize=15cm
\epsffile}
\newcommand{\efi}{\end{figure}}
\newcommand{\efig}{\end{figure*}}
\newcommand{\no}{\nonumber}

\title[Dark siren distance calibration with LSS]{Calibrating systematic errors in the distance
  determination with the  luminosity-distance space large scale
  structure of dark sirens and 
  its potential applications}
\author[Zhang \& Yu]{Pengjie Zhang$^{1,2,3}$, Hai Yu$^{1,2}$\\
$^1$Department of Astronomy, School of Physics and Astronomy, Shanghai
Jiao Tong University, Shanghai, 200240, China\\ 
$^2$Key Laboratory for Particle Astrophysics and Cosmology
(MOE)/Shanghai Key Laboratory for Particle Physics and Cosmology,
China\\
$^3$Tsung-Dao Lee Institute, Shanghai Jiao Tong University , Shanghai
200240, China\\
Email to: zhangpj@sjtu.edu.cn; yu\_hai@sjtu.edu.cn
}
\begin{document}
\maketitle
\begin{abstract}
The cosmological luminosity-distance can be measured from 
gravitational wave (GW) standard sirens, free of
astronomical distance ladders and the associated
systematics. However, it may still contain systematics arising from various
astrophysical, cosmological and experimental sources. With the large
amount of  dark standard sirens of upcoming third generation GW
experiments, such potential systematic bias can be diagnosed and corrected by statistical tools of  the
large scale structure of the universe. We estimate that, by
cross-correlating the dark siren luminosity-distance space
distribution and galaxy redshift space distribution, multiplicative error $m$
in the luminosity distance measurement can be constrained with
$1\sigma$ uncertainty $\sigma_m\sim
0.1$. This is already able to distinguish some binary black hole
origin scenarios unambiguously. Significantly better constraints and
therefore more
applications may be achieved by more advanced GW experiments.
\end{abstract}
%%%%%%%%%%%%%%%%%%%%%%%%%%%%%%%%%%%%%%%%%%%%%%%
%       Introduction
%%%%%%%%%%%%%%%%%%%%%%%%%%%%%%%%%%%%%%%%%%%%%%%

%\section{Introduction} 
\section{Introduction}
About 50 gravitational wave (GW) events of binary compact star coalescence have been detected by
advanced LIGO and advanced Virgo collaboration (LVC)\citep{2016PhRvL.116f1102A,2017PhRvL.118v1101A,2017PhRvL.119n1101A,2017PhRvL.119p1101A,2019PhRvX...9c1040A,
  2020ApJ...892L...3A,2020ApJ...896L..44A,2020arXiv200408342T,Abbott2020arXiv201014527A,Abbott2020arXiv201014533T}. The
total number of 
detections will b dramatically increased by three orders of magnitude by the third generation GW
experiments such as the proposed Einstein Telescope
(ET\footnote{https://www.et-gw.eu/}) and Cosmic Explorer
(CE, \citet{2015PhRvD..91h2001D,2017CQGra..34d4001A,CosmicExplorer}). These
experiments will be able to detect
binary black holes (BBHs) within our horizon
\citep{2015PhRvD..91h2001D}. Given the estimated 
BBH merger rate $R=53.2^{+58.5}_{-28.8}$ Gpc$^{-3}$ yr$^{-1}$ ($90\%$
credibility, \citet{2019ApJ...882L..24A}),  $\sim 10^5$ BBH detections per year are
achievable.  The angular localization accuracy
will improve to $\la 1$ deg$^2$, and the measurement accuracy of
luminosity-distance through the standard siren method\citep{1986Natur.323..310S, 2017Natur.551...85A}  will reach $\sigma_D/D\la 0.1$
\citep{Zhao18}. The detection of higher spherical harmonics beyond the quadruple mode
(GW190814, \citet{2020ApJ...896L..44A} \& GW190412, \citet{2020arXiv200408342T}) and the analysis using
the full waveforms instead of the restricted waveforms\citep{2007PhRvD..76j4016A,2007CQGra..24.1089V,2009PhRvD..79h4032A,2019arXiv191004565R} will lead to
further significant improvement in the angular resolution and distance
determination.

Binary neutron star (BNS) GW events have weaker amplitude, so the redshift
coverage,  angular resolution and distance determination accuracy by ET and CE
will be significantly 
worse than that of BBH GW events \citep{Zhao18,CosmicExplorer}. Nevertheless, more ambitious GW 
experiments such as Big Bang Observer (BBO,
\citet{2006PhRvD..73d2001C})  are capable of achieving $1\%$ distance
measurement and arcminute angular localization. Since BNS event rate
is an order of magnitude higher
\citep{2017PhRvL.119p1101A,2019PhRvX...9c1040A,2020ApJ...892L...3A},
eventually we would expect $\sim 10^6$ GW events (mostly BNS events)
per year of known 3D positions all over the observable universe.

These upcoming 3D mappings of the universe through GW events will enable
the measurement and application of a new type of large scale structure
(LSS) of the universe, namely the 
luminosity-distance space LSS \citep{Zhang18a,Zhang18b,2020arXiv200704359N,2020arXiv200706905L}.  Most of these
GW events will not have electromagnetic (EM) counterparts or host
galaxies identified, so the usual standard siren cosmology techniques
\citep{1986Natur.323..310S,2005ApJ...629...15H,2009PhRvD..80j4009C,2017Natur.551...85A,2019ApJ...876L...7S,2019arXiv190806060T}
are not
applicable. Nevertheless, these dark standard sirens construct an excellent data set of LSS.
Due to the fact that  gravitational waves are transparent while EM
waves are not, the observed GW event spatial distribution is immune to 
galactic and extragalactic contaminations. Its spatial selection
function is expected to be more
homogeneous and well understood. Together with the potentially full coverage
of cosmic volume within the horizon, the luminosity-distance space LSS
is promising to provide clean information on all linear LSS modes
of our universe. Furthermore, this LSS is spatially correlated with
the usual LSS in redshift space and various 2D projections (e.g. weak
lensing), making them highly complementary \citep{Zhang18a,Zhang18b,2016PhRvL.116l1302N,2016PhRvD..93h3511O,2018PhRvD..98b3502N,2018arXiv180806615M, 2019ApJ...876L...7S, 2019arXiv190806060T}. 

The dark siren based LSS does not require EM identification of EM
counterparts or host galaxies. However, all promising 
applications of the dark siren based  LSS are built on unbiased
measurement of the luminosity-distance $D^{\rm obs}_L$, with respect to the
cosmological luminosity-distance $D_L^{\rm cos}$. Various
experimental, astrophysical, and cosmological issues may cause
systematic bias in the distance 
measurement, analogous to redshift measurement in EM observations. (1)
The most obvious one is  calibration error in the GW
strain amplitude (e.g. \cite{2020arXiv200502531S}). Although it is expected  to be small
(several percent for current experiments), independent verification is
useful.  Calibration errors also affect the phase and therefore the
inferred distance. (2) One example of cosmological
issues is modification to general relativity, which may affect the GW
propagation and/or the GW generation
(e.g. \cite{2007ApJ...668L.143D,2018PhRvD..97j4066B,2018FrASS...5...44E}). If 
graviton has 
mass or if it can propagate in extra dimensions, 
the resulting $D_L^{\rm obs}$ will be larger than the corresponding
luminosity-distance of EM counterpart, which equals to the
luminosity-distance ($D^{\rm cos}$) of the FRW universe.  (3) One example of
astrophysical issues is related to the environment of GW
events. This is motivated by the challenge in explaining the origin of high mass BBH ($\ga
20M_\odot$) by stellar evolution models.  \cite{2019MNRAS.485L.141C} argued that the
observed ``high mass'' BBHs may preferentially reside within $\la 10$
Schwarzschild radius of the supermassive black hole of host
galaxies. These BBH events then suffer an extra gravitational redshift
$z_{\rm gra}$  and an extra Doppler redshift $z_{\rm Dop}$.  The inferred black hole
masses will be overestimated by $(1+z_{\rm gra})(1+z_{\rm Dop})$. In this scenario,
the inferred luminosity-distance is overestimated by the same factor,
$D_L^{\rm obs}=D_L^{\rm cos}(1+z_{\rm gra})(1+z_{\rm Dop})$.  This effect can lead to a factor of $\sim 2$
bias in $D_L$.  Other BBH environment effects may also exist. Gas
around the BBH orbit affects the orbital  motion and therefore affects the GW waveform
\citep{2019arXiv190611055C}. Although this effect is likely negligible for
high frequency observations such as LIGO, ET and CE, this demonstrates
the need to comprehensive investigation of BBH environment and its
impact on the $D_L$ determination. 

Given the possible existence of significant systematic error in $D_L$,
standard siren cosmology will be significantly impacted. It is the
purpose of this paper to provide a model-independent method of
diagnosing and correcting such systematics. Significant systematic error in $D_L$
would cause abnormal behavior
in the angular correlation of GW dark sirens and galaxy
distributions.  Similar behavior is well known in the context of
LSS. One early application is the SZ tomography \citep{Zhang01,Shao11a}, where
cross-correlating wit galaxies help recover the redshift information
of the thermal Sunyaev Zel'dovich (SZ) effect. It is also widely used
in the cross calibration \citep{2008ApJ...684...88N} and the
self-calibration of galaxy photometric redshift errors\citep{2006ApJ...651...14S,Zhang10a,2017ApJ...848...44Z}. Similar
techniques have been designed to identification of dark siren redshifts
\citep{2016PhRvL.116l1302N,2016PhRvD..93h3511O,2018PhRvD..98b3502N,2018arXiv180806615M,Zhang18b}.
The question is to convert the directly measured cross
correlation into unbiased constraint on systematic error in $D_L$. We
find that this problem is essentially identical to identify the true
redshift distribution of GW dark sirens. For the later purpose, we
have proposed the $E_z$ estimator\citep{Zhang18b}. It is essentially the same
estimator that one of the authors proposed in the context of SZ
tomography \citep{Zhang01}.  \citet{Zhang18b} proved that, it is unbiased in
recovering the true redshift distribution, insensitive to model
assumptions on dark siren and galaxy distribution. The same $E_z$ estimator,
when applying to the current problem, will not only diagnose the
existence of  systematic bias in $D_L$, but also constrain it
unbiasedly. 

This paper is organized as follows. In \S \ref{sec:method}, we 
briefly describe the $E_z$ estimator in the context of calibrating systematic
error in $D_L^{\rm obs}$. In \S \ref{sec:result} we estimate its
constraining power. In term of the multiplicative error $m$, we find
that ET and CE, together with pre-existing galaxy surveys,  will be able to constrain with
$\sigma_m\sim 0.1$.  We discuss and summarize in \S
\ref{sec:conclusion}. 

\section{Cross-calibration with the $E_z$ estimator}
\label{sec:method}
Any error leading to $\langle D_L^{\rm obs}-D_L^{\rm
  cos}\rangle \neq 0$ belongs to systematic error. Following the common
parametrization in cosmic shear measurements \citep{STEP1,STEP2}, we
can classify them into multiplicative errors and additive errors,
\ba
D_L^{\rm obs}=D^{\rm cos}_L(1+m)+\epsilon\ .
\ea
$\epsilon$ is the additive error. When $\langle \epsilon\rangle \neq 0$
or when $\epsilon$ has spatial correlation (analogous to the galaxy
intrinsic alignment in cosmic shear), it is a systematic
error. Otherwise it will be  the statistical
error in the distance determination.  For the current paper, we will
only consider the later case, as would be caused by instrumental noise of GW experiments. Namely we consider $\epsilon$ as a
random variable with zero mean and pdf $p(\epsilon)$. For brevity, we
assume $p(\epsilon)$ to be Gaussian, with $\sigma_\epsilon=\sigma_{
  D_L^{\rm obs}}$. 
$m$ is the multiplicative error in
the luminosity-distance determination, and $\langle m\rangle\neq 0$.  As discussed in the previous section, it can
have various origins. For the astrophysical origin proposed by
\citet{2019MNRAS.485L.141C}, $1+m=(1+z_{\rm gra})(1+z_{\rm
  Dop})$. \citet{2019MNRAS.485L.141C} argued that for BBHs with mass
$\ga 20M_{\odot}$, $m\sim 1$. But BBHs with lower mass may reside in
different environments and are free of this bias. Therefore later we
will consider two fiducial cases of $m=0$ and $m=1$.

With the presence of $m\neq 0$,  the true redshift distribution of
dark sirens within a distance bin ($D_1<D^{\rm obs}_L<D_2$) is 
\ba
\label{eqn:ngw}
\bar{n}_{\rm GW}^{\rm cos}(z)&=&\int_{D_1}^{D_2} dD_L^{\rm obs}
\int_0^\infty dD^{\rm cos}_L \delta_D\left(z-z(D^{\rm cos}_L)\right)\no\\
&&\times 
p(D^{\rm cos}_L|D_L^{\rm obs}) n_{\rm
  GW}^{\rm obs}(D_L^{\rm obs})\ .
\ea
Here $n_{\rm
  GW}^{\rm obs}(D_L^{\rm obs})$ is the number of dark sirens per unit
$D_L^{\rm obs}$ interval. $p(D^{\rm cos}_L|D_L^{\rm obs})$ is the PDF
of $D^{\rm cos}_L$ given $D_L^{\rm obs}$. It is determined by the statistical error ($\epsilon$)
probability distribution $p(\epsilon)$ and $m$, where $\epsilon=D_L^{\rm obs}-D^{\rm
  cos}_L(1+m)$. Fluctuations in the dark siren surface density
is then
\ba
\delta_{{\rm GW}}^{\Sigma}(\hat{\theta})&=&\bar{\Sigma}_{{\rm GW}}^{-1}\left[\int_{0}^{\infty} \delta_{\rm
  GW}(z,\hat{\theta}) \bar{n}^{\rm cos}_{{\rm GW}}(z) dz\right]\ .
\ea
Here the mean surface number density 
\ba
\bar{\Sigma}_{{\rm GW}}&=&\int_{D_1}^{D_2}
\bar{n}_{\rm GW}^{\rm obs}(D_L^{\rm obs}) dD_L^{\rm obs}=\int_0^\infty
\bar{n}^{\rm cos}_{{\rm GW}}(z)dz\ .
\ea
$\delta_{\rm GW}(z,\hat{\theta})$ is the number overdensity of dark sirens at
cosmological redshift $z$ and angular direction $\hat{\theta}$.  

These dark sirens have angular cross-correlation with galaxies. We
have the freedom to weigh the galaxy number distribution $n_g(z)$ with
an arbitrary weighting function $W_g(z)$. For the weighted galaxy
distribution,
 \ba
\delta_{\rm g}^{\Sigma}(\hat{\theta})&=&\bar{\Sigma}^{-1}_{\rm
    g}\left[\int_{0}^{\infty} \delta_{\rm g}(z,\hat{\theta})
  \bar{n}_{\rm g}(z)W_g(z) dz\right] \ .
\ea
Here the mean galaxy surface number density $\bar{\Sigma}_{\rm g,j}=\int_{0}^{\infty}  \bar{n}_{\rm
  g}(z)W_g(z)dz$.   The GW-galaxy cross and galaxy auto angular power
spectra at multipole $\ell$ are 
\ba
C^{\rm GW-g}(\ell)&=&\frac{\int P_{\rm
    GW-g}(k,z)\bar{n}^{\rm cos}_{\rm GW}\bar{n}_{\rm
    g}W_g\chi^{-2}\frac{dz}{d\chi}dz}{\bar{\Sigma}_{\rm GW}\bar{\Sigma}_{\rm
    g}}\ , \\
C^{\rm g}(\ell)&=&\frac{\int P_{\rm g}(k,z)(\bar{n}_{\rm
    g}W_g)^2\chi^{-2}\frac{dz}{d\chi}dz}{\bar{\Sigma}_{\rm
    g}^2}\ . 
\ea
Here $\chi$ is the comoving radial distance. $P_{\rm g}(k,z)$ and $P_{\rm GW-g}(k,z)$ are the 3D galaxy and
galaxy-GW host galaxy power spectrum respectively, with
$k=\ell/\chi(z)$ in the above integrals.
The above expressions adopt a flat universe cosmology. But the
methodology holds for curved universe as well.   

The true redshift distribution (or $m$) of GW dark standard sirens can be
obtained by maximizing the $E_z$ estimator \citep{Zhang18b}, 
\ba
E_z(\ell|W_g)=\frac{C_{\rm GW-g}(\ell|W_g)}{\sqrt{C_g(\ell|W_g)}}\ .
\ea
Here we highlight the dependence of $E_z$, $C_{\rm GW-g}$ and $C_g$ on
the weighting function $W_{\rm g}$. By varying $W_g$ to find the maximum of
$E_z$, unbiased estimation of the true redshift distribution of dark
sirens can be obtained (refer to \citet{Zhang18b} for the proof).
The
reason is that $E_z$ reaches maximum when the cross-correlation
coefficient between surface overdensities of GW and weighted
galaxies reaches
maximum. This is naturally achieved when the two distributions match each
other in redshift. For
$W_g$ of free functional form, the numerical variation can
be challenging. However, in the context of the current paper, the
situation is simpler and we can take a constrained form of $W_{\rm g}$. Since $m$ is the only unknown parameter, the
form of $W_g$ can be  fixed up to an unknown
fitting parameter  $m^{\rm fit}$ (namely replacing $m$ in
Eq. \ref{eqn:ngw} with $m^{\rm fit}$),
\ba
\label{eqn:Wgfit}
W_{\rm g}(z|m^{\rm fit})=\frac{\bar{n}^{\rm cos}_{\rm GW}(z|m^{\rm fit})}{\bar{n}_g(z)}\ .
\ea
When $m^{\rm fit}=m^{\rm true}$,  the weighted galaxy redshift
distribution $\bar{n}_{\rm g}W_{\rm g}\propto \bar{n}^{\rm cos}_{\rm
  GW}$. Namely the weighted galaxy redshift distribution will match
that of the GW redshift distribution exactly. This will result in the
maximum $E_z$. So the form of Eq. \ref{eqn:Wgfit} includes returns
unbiased estimation of $m$. This is also numerically shown in
Fig. \ref{fig:Ez}. Therefore we will adopt this
constrained form of $W_{\rm g}$.

For brevity, hereafter we neglect the superscript ``fit'' in $m^{\rm
  fit}$ where it does not cause confusion with the fiducial value
$m^{\rm fid/true}$. 
The bestfit value of $m$ is obtained when $\partial E_z/\partial m=0$. Then we take
the data as $\partial E_z/\partial m$. The posterior probability of $m$ is 
\ba
p(m)&\propto& \exp\left(-\frac{1}{2}\Delta \chi^2\right)\ , \no\\
\Delta \chi^2&=&\sum_\ell\left(\frac{\partial E_z}{\partial m}
  C^{-1}\frac{\partial E_z}{\partial m}\right)\ .
\ea
Here $C$ is the covariance matrix of $\partial E_z/\partial m$ at
multipole $\ell$, whose full expression
is given in \cite{Zhang18b}. Statistical errors in $\partial
E_z/\partial m$ of different $\ell$ do not correlate, so we can sum
over contributions from each $\ell$ bin to obtain the total $\Delta
\chi^2$. 

\bfi{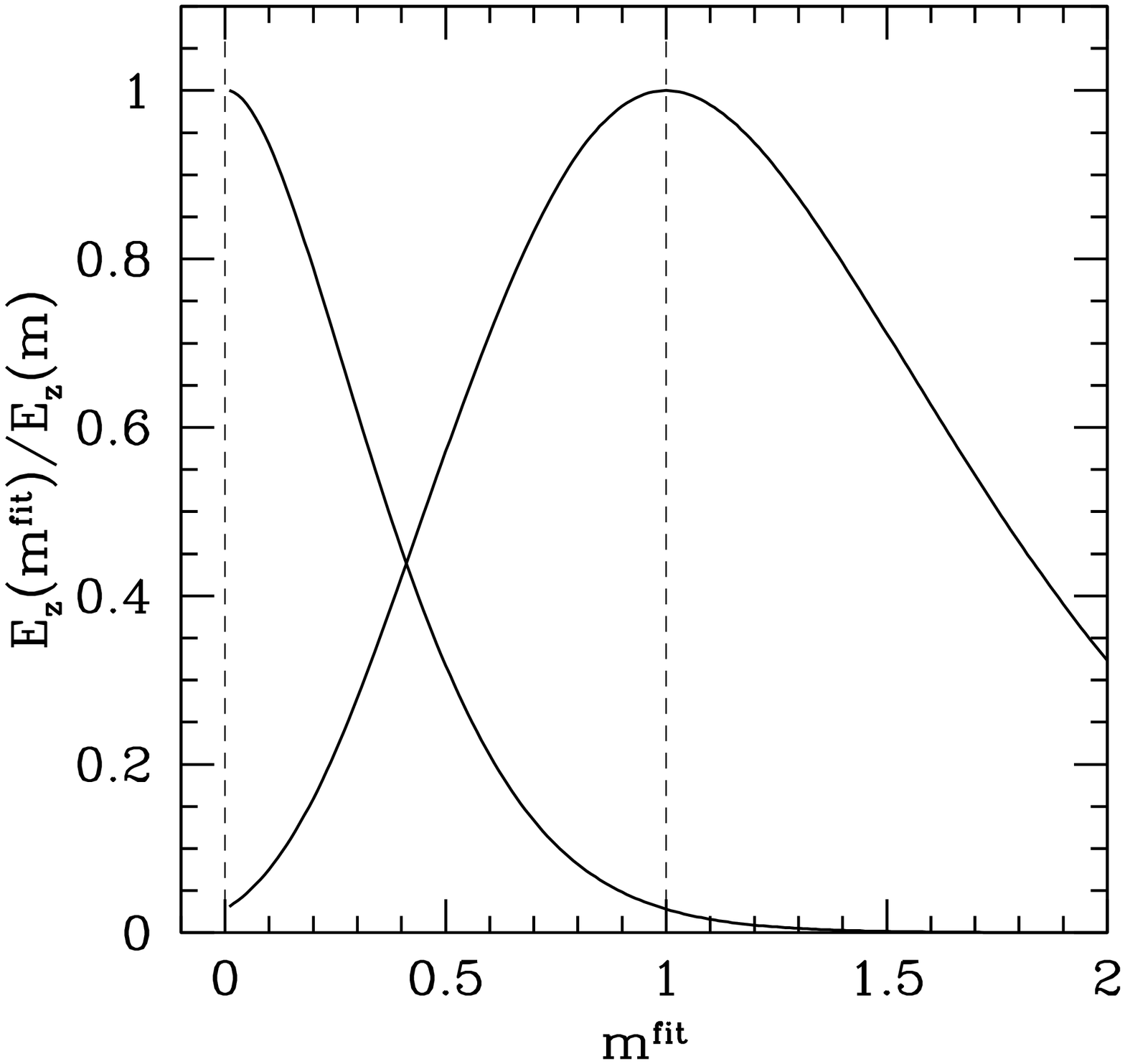}
\caption{$E_z$ as a function of the fitting parameter $m^{\rm fit}$, for the luminosity bin $D_LH_0/c\in
(1.19,1.98)$. The
  fiducial $m$ is adopted as $m=0$ and $m=1$, respectively. Indeed, $E_z(m^{\rm
    fit})$ peaks at $m^{\rm fit}=m$. Namely it provides unbiased
  estimation of $m$. We have normalized its amplitude with
  $E_z(m^{\rm fit}=m)$. \label{fig:Ez}}
\efi
\section{Error forecast}
\label{sec:result}
We apply the $E_z$ estimator to evaluate its performance in
constraining $m$, for third generation GW experiments such as ET and
CE.  We adopt the fiducial cosmology as the $\Lambda$CDM cosmology with
$\Omega_m=0.268$, $\Omega_\Lambda=1-\Omega_m$, $\Omega_b=0.044$,
$h=0.71$, $\sigma_8=0.83$ and $n_s=0.96$. We consider the fiducial $m=1$, as would be expected in the
scenario of \citep{2019MNRAS.485L.141C}.  The first question is that
whether the data alone allows model-independent discrimination against
the case of $m=0$. If yes, then we need to verify that the bestfit
indeed gives $m=1$ and need to quantify $\sigma_m$. 

Implementing the $E_z$ method requires deep and wide galaxy surveys
with accurate redshift distribution. By the time of these GW
experiments, stage IV spectroscopic redshifts surveys of $\sim
10^{8}$ galaxies such as
CSST \citep{2018cosp...42E3821Z,2019ApJ...883..203G},
DESI \citep{DESI16}, Euclid \citep{Euclid}, PFS
\citep{2014PASJ...66R...1T}, WFIRST
\citep{WFIRST15}, and intensity mapping surveys
(e.g. SKA\citep{2015aska.confE..19S} and SPHEREx
\citep{2014arXiv1412.4872D,2018cosp...42E.875D},) will likely be
completed and satisfy the need. More advanced surveys
with $\sim 10^9$ galaxies with spectroscopic redshifts and
deeper redshift coverage may also be available
(e.g. \citet{2015aska.confE..17A,Cosmicvision,2019BAAS...51c.508W,2019BAAS...51c..72F}). We also expect photometric redshift of imaging
surveys  such as LSST will improve to the required accuracy for
implementing the $E_z$ method. Therefore for a bin width of $\Delta
z\sim 0.2$, we expect $N_g$ of the order $10^{7-8}$.

For the GW experiments, we target at third generation experiments like
ET and CE. Unlike more advanced GW experiments of
percent-level accuracy in luminosity distance determination and
arcminute angular resolution for both BBHs and BNSs \citep{2006PhRvD..73d2001C}, these
experiments may only reach $\sim 1$deg$^2$ in the
angular resolution for BBHs \citep{Zhao18}. This results into a
suppression of $C_{\rm GW-g}(\ell)$ significant at $\ell \ga 100$,
\ba
C_{\rm GW-g}(\ell)&\rightarrow& C_{\rm GW-g}(\ell)S(\ell)\ ,\no \\
E_z(\ell)&\rightarrow& E_z(\ell)S(\ell)\ , \no \\
\frac{\partial E_z}{\partial m}&\rightarrow& \frac{\partial
  E_z}{\partial m}S(\ell)\ .
\ea
Here $S(\ell)$ is the suppression caused by uncertainties in the
angular localization. We take a Gaussian form
$S(\ell)=\exp(-\ell^2\sigma^2_\theta/2)$, whose effective area is
$2\pi\sigma_\theta^2$. So the typical $\sigma_\theta\sim
1^{\circ}=\pi/180$ for ET and CT.

These experiments have a typical distance error $\sigma_D/D\la 0.1$
\citep{Zhao18}. Therefore luminosity bin width smaller than this
is 
unnecessary. Furthermore, the bin width needs to be sufficiently large
to include enough GW events. We then choose the luminosity bins
$D_LH_0/c\in (0.51,1.19)$, $(1.19,1.98)$, $(1.98,2.83)$. They
correspond to   $z\in (0.4,0.8)$, $(0.8,1.2)$, $(1.2,1.6)$, if the
distance measurement is exact. The analysis can be carried out for
other choices of luminosity-distance  bins. The investigated ones,
with bin width much larger than $\sigma_D$, are chosen mainly for the
purpose of sufficient GW events. The number of GW dark sirens in each bin is $\sim
10^4$, with too large uncertainty to predict reliably. Therefore we
adopt $N_{\rm GW}=10^4$ as the fiducial value. Associated errors can
be conveniently rescaled by $N^{-1/2}_{\rm GW}$ to other values. The
bottom line is that we will work under the situation $N_{\rm GW}\sim
10^4\ll N_{\rm g}\sim 10^{7-8}$.

Fig. \ref{fig:Ez} shows the dependence of $E_z$ on the fitting
parameter $m^{\rm fit}$, for the luminosity bin $D_LH_0/c\in
(1.19,1.98)$. We show the cases of two fiducial $m=0.0,1.0$
respectively. It is clear that $E_z$ reaches the maximum when $m^{\rm
  fit}=m$. This verifies the theoretical prediction that the $E_z$
estimator is unbiased. Furthermore, for the same $m^{\rm fit}$, $E_z$
of different fiducial $m=0,1$ shows significant difference. 

\bfi{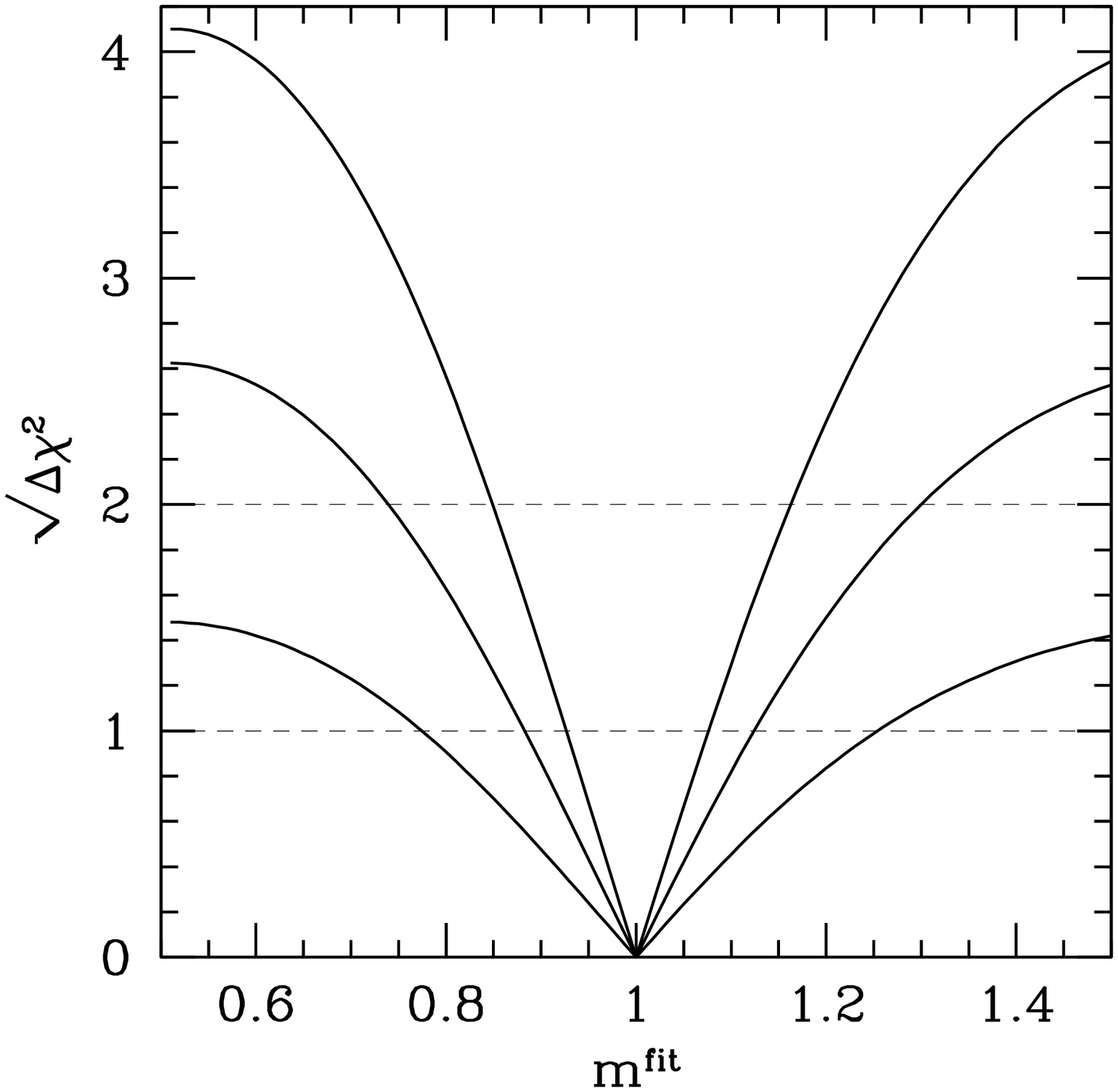}
\caption{$\sqrt{\Delta \chi^2}$ as a function of $m^{\rm fit}$, which
  determines the posterior probability of $m$ ($p(m^{\rm fit})\propto
  \exp(-\Delta \chi^2/2)$). The fiducial $m=1$. The luminosity distance range is
  $(1.19,1.98) c/H_0$. We adopt $N_{\rm GW}=10^4$, $b_{\rm GW}=1$,
  $\sigma_{\ln D}=0.1$. From top to bottom,
  $\sigma_\theta=0.5^{\circ},1^{\circ}, 2^{\circ}$ are adopted
  respectively. The $1\sigma$ error in $m$ ranges from $0.07$ to
  $0.24$.  \label{fig:chi2}}
\efi

The constraining power of $E_z$ on $m$ is then quantified by the
measurement errors in $E_z$ and $\partial E_z/\partial m$.  The measurement error in $E_z$ arises both from that in $C_{\rm GW-g}$
and $C_{\rm g}$. Given the availability of powerful galaxy surveys listed
above, shot noise in $C_g$ at scales of interest ($\ell\sim
100$) is negligible. Namely $C_{\rm g,N}=4\pi f_{\rm sky}/N_{\rm g}\ll
C_{\rm g}$. But that in the GW dark siren distribution
dominates over the cosmic variance. Namely the shot noise power
spectrum $C_{\rm GW,N}=4\pi f_{\rm sky}/N_{\rm GW}\gg C_{\rm
  GW}$. Therefore the statistical error $\sigma_{E_z}$ in $E_z$ of a
multipole bin centered at $\ell$ and with bin width $\Delta \ell$ is dominated by error
in $C_{\rm GW-g}$. We then have
\ba
\left(\frac{E_z}{\sigma_{E_z}}\right)^2_\ell&\simeq& \frac{2\ell \Delta \ell f_{\rm
  sky}}{1+(C_{\rm GW}+C_{\rm GW,N}) (C_{\rm g}+C_{\rm
      g,N})/C_{\rm GW-g}^2} \ . \no
\ea
It can be further simplified, 
\ba
\left(\frac{E_z}{\sigma_{E_z}}\right)^2_\ell&\simeq &\frac{2\ell \Delta \ell f_{\rm
  sky}}{1+C_{\rm GW,N}C_{\rm g}/C_{\rm GW-g}^2} \no \\
&\simeq& \frac{2\ell \Delta \ell f_{\rm
  sky}}{C_{\rm GW,N}C_{\rm g}/C_{\rm GW-g}^2}=\frac{\ell \Delta\ell}{2\pi}N_{\rm GW} E_z^2\no \\
&\simeq & 16 \left(\frac{\ell \Delta \ell}{10^4}\right)
\left(\frac{N_{\rm GW}}{10^4}\right) \left(\frac{E_z}{10^{-3}}\right)^2\ . 
\ea
Notice that $N_{\rm GW}$ is the total number of GW events in the sky
area of the corresponding galaxy survey, which only covers a fraction
$f_{\rm sky}$ of the whole sky.

The
statistical error in $\partial E_z/\partial m$ is $\sigma_{\partial
  E_z/\partial m}=C^{1/2}$.  The full expression of
$C$ is given in \cite{Zhang18b} . Adopting the same approximation in
$E_z$, we  have 
\ba
C&\simeq& \frac{1}{2\ell \Delta \ell f_{\rm sky}}\frac{4\pi f_{\rm
    sky}}{N_{\rm GW}} \no \\
&\times& \frac{\int W^2_{g,m}P_g\bar{n}_g^2\chi^{-2}(dz/d\chi)dz}{\int
  W^2_{g}P_g\bar{n}_g^2\chi^{-2}(dz/d\chi)dz}\ \no\\
&=& \frac{2\pi}{\ell \Delta \ell N_{\rm GW}} \frac{\int W^2_{g,m}P_g\bar{n}_g^2\chi^{-2}(dz/d\chi)dz}{\int
  W^2_{g}P_g\bar{n}_g^2\chi^{-2}(dz/d\chi)dz}\ .
\ea
 Here $W_{g,m}(z,m)\equiv \partial W_g(z,m)/\partial m$.

Fig. \ref{fig:chi2} shows the resulting $\Delta \chi^2$ as a function
of the fitting parameter $m^{\rm fit}$, for the fiducial case of
$m=1.0$ and the luminosity bin $D_LH_0/c\in
(1.19,1.98)$. We adopt $\sigma_D/D=0.1$ and show its dependence on the
angular resolution $\sigma_\theta$. $\Delta \chi^2=1$ determines the $1\sigma$ error
$\sigma_m$ in the $m$ constraint. The resulting $\sigma_m$ is shown in
Table \ref{table:m}. We show its dependence on the luminosity bin,
angular resolution and $\sigma_D/D$. 

In general, $\sigma_m\sim 0.1$ is
achievable. This will robustly distinguish the  scenario of BBH
origin with $m=1$ proposed in \cite{2019MNRAS.485L.141C} from the
commonly assumed $m=0$ case. To further check these scenarios, we can split BBHs of
the same luminosity bin into two samples of mass above and below
$20M_\odot$, and apply the $E_z$ estimator separately. If the scenario
proposed by \cite{2019MNRAS.485L.141C} is valid, we will find
that $m\sim 1$ and $\sigma_m\sim 0.1$ for the sample of $M>20
M_\odot$,  while $m\sim 0$ and $\sigma_m\sim 0.1$ for the sample of $M<20
M_\odot$. If we apply the $E_z$ estimator to the whole sample, we will
find that $m$ significantly deviates from zero. Furthermore, the
minimum $\Delta \chi^2$ will be significantly larger than unity,
meaning bad fit of the data with a single $m$ and multiple origins of
BBHs. 

%%%%%%%%%%%%%%%%%
\begin{table}
	\caption{$\sigma_m$, the constraining power in $m$, as a
          function of observed luminosity distance range, angular
          location accuracy $\sigma_\theta$, and distance measurement error
          $\sigma_{\ln D}$. The fiducial multiplicative error
          parameter $m=1$. The listed luminosity distance ranges
          correspond to redshift ranges $(0.4,0.8)$, $(0.8,1.2)$ and
          $(1.2,1.6)$ respectively, {\it if we interpret them assuming
          $m=0$}. $\sigma_\theta$ is the angular resolution, and third generation GW
          experiments have $\sigma_\theta\sim 1^{\circ}$. The quoted values of
          $\sigma_m$ outside
          of the parentheses have $\sigma_{\ln D}=0.1$ and those
          inside have $\sigma_{\ln D}=0.2$. The
          number of dark sirens within the observed luminosity-distance range is adopted as $N_{\rm
            GW}=10^4$.  } 
	\label{table:m}
	\begin{center}
		\scriptsize
		\begin{tabular}{@{}c|c|c|c}
			\hline
		$D_L^{\rm obs}H_0/c$	&
                           $\sigma_\theta=0.5^{\circ}$&$\sigma_\theta=1^{\circ}$
                  & $\sigma_\theta=2^{\circ}$\\ 
\hline
              $(0.51,1.19)$& $\sigma_m=0.07(0.11)$ & $0.11(0.16)$ & $0.19(0.26)$\\
%			\hline
	      $(1.19,1.98)$	 &$\sigma_m= 0.07(0.12)$ &$0.12(0.20)$ &$0.24(0.40)$\\
%			\hline 
              $(1.98,2.83)$       &$\sigma_m=0.08(0.15)$  & $0.13(0.27)$ &
                                                                  $0.35(--)$\\ 
\hline
		\end{tabular}
	\end{center}
\end{table}
%%%%%%%%%%%%%%%%%%%%%%%%%%%%%%

There are many factors affecting the constraining power on
$m$. Since
$C\propto N_{\rm GW}^{-1}$ and $dE_z/dm\propto b_{\rm GW}$, we have
\ba
\sigma_m\propto N_{\rm GW}^{-1/2}b^{-1}_{\rm GW}\ .
\ea
Here $b_{\rm GW}$ is the bias of GW dark sirens. We adopt $b_{\rm
  GW}=1$ and $N_{\rm GW}=10^4$ as the fiducial 
values. We remind that, although $\sigma_m$ depends on $b_{\rm GW}$,
the bestfit $m$ does not \citep{Zhang18b}. Furthermore, $\sigma_m$ does
not depend on the galaxy bias $b_g$, under the limit of negligible shot
noise in galaxy distribution.  The major limiting factor, from the
observational side, is $N_{\rm GW}$.  Therefore the constraining power
on $m$ will also significantly depend on the S/N threshold of GW events, which changes
$N_{\rm GW}$ significantly. 

The dependences on other factors are more complicated. These include the angular
localization uncertainty $\sigma_\theta$, the distance measurement
error $\sigma_{\ln D}$, the range of luminosity distance and the true
value of $m$. One significant  dependence is on $\sigma_\theta$
(Fig. \ref{fig:chi2} \& Table \ref{table:m}). $\sigma_\theta$ sets the
maximum $\ell_{\rm max}\sim 2/\sigma_{\theta}\simeq
115/(\sigma_{\theta}/1^{\circ})$, available to the given GW
experiment. This factor alone would leads to $\sigma_m\propto
\sigma_\theta$. But since shot noise also increases with increasing
$\ell$, the dependence on $\sigma_\theta$ is slightly weaker.  The
dependence of $\sigma_m$ on $\sigma_D/D$ is also significant, but for
a different reason. Larger $\sigma_D/D$ means wider (true) redshift
distribution in GW dark sirens, making the effort of constraining $m$ more
difficult. $\sigma_m$ increases slowly with increasing
redshift/distance (Table \ref{table:m}), for fixed $\sigma_\theta$,
$\sigma_{\ln D}$, and $b_{\rm GW}$. The reason is that same $\ell$
means smaller $k$ at higher redshift/distance, and therefore weaker
spatial correlation signal.

\section{Discussion and conclusion}
\label{sec:conclusion}
%\section{Acknowledgement}
We have investigated the possibility of calibrating multiplicative
errors in the dark standard siren distance determination. Such errors cause
systematic deviation of the measured luminosity distance from the true cosmological
luminosity distance.  Through cross correlation with pre-existing
galaxy redshift surveys, such systematic error can be constrained and
the constraint can be made model-independent and unbiased by the
proposed $E_z$ estimator. The constraining power is mainly limited by
the angular resolution and distance determination uncertainty. For
ET/CE-like GW experiments, $\sigma_m\sim 0.1$ is achievable.  This
constraint will provide unambiguous distinction of some BBH origin
mechanisms (e.g. \cite{2019MNRAS.485L.141C}). But we caution that
  the discriminating power relies on the capability to appropriately
  select GW samples. If the selected GW sample is composed of  GW events of mixed $m$,
  we may only constrain the mean $m$, but lose the power to idendify
  those sub-samples with significant $m\neq 0$. 

Multiplicate error is not the only form of systematic errors in the
distance determination. Additive error with non-zero average ($\langle
\epsilon\rangle\neq 0$) shifts the mean value of $D_L^{\rm obs}$,
analogous to $z_{\rm bias}$ in photo-z errors. Weak lensing
magnification in the luminosity-distance measurement and
magnification bias in the number distribution of GW events and
galaxies, can also bias the inference of GW redshifts. Given the
weak signal, the impact of weak lensing on $m$
determination is negligible for $\sigma_m\sim 0.1$. But it may become
non-nelgigible when $\sigma_m$ approaches $1\%$ or below.  Other more complicated
form of systematic errors may also exist. Nevertheless, the same dark
siren LSS based calibration technique and the specific $E_z$ estimator
are applicable.

The expected $\sigma_m\sim 0.1$ is for third generation GW
experiments and the main limiting factors are the angular resolution
and the distance determination statistical errors. Although it is already useful for certain applications,
it  does not meet the requirement to diagnose calibration error in GW
strain measurements, or modification of GW $D_L$ and EM $D_L$ by
modified gravity models still surviving. These applications require
$\sigma_m\sim 0.01$. Such constraining accuracy can be achieved by
more advanced surveys such as BBO. Its arcminute resolution means that
multipole modes with $\ell\ga 10^{2-3}$ are accessible. Its percent
level distance determination accuracy makes a finer luminosity bin
size useful. Its sensitivity to BNS events within the horizon will
increase the dark siren number density by an order of magnitude,
making $\sim 10^5$ GW events per luminosity-distance bin 
feasible. With these threefold improvements, $\sigma_m=0.01$ is promising and new applications
are possible. 

\section*{Data availability}
No new data were generated or analyzed in support of this research. 

{\bf Acknowledgement}.---
This work was supported by the National Science Foundation of China
11621303 \& 11653003. We thank Xian Chen for valuable information on
the BBH origin mechanisms. 
\bibliographystyle{mnras}
\bibliography{mybib}
\end{document}